\documentclass[reprint,twocolumn,superscriptaddress,showpacs,preprintnumbers,amsmath,amssymb,aps, pra]{revtex4}

\usepackage{amssymb}
\usepackage{amsfonts}
\usepackage{amsmath}
\usepackage{cases}
\usepackage{stmaryrd}\usepackage{tipa}
\usepackage[dvips]{graphicx}
\usepackage{dcolumn}
\usepackage{bbold}
\usepackage{graphicx}
\usepackage{subfigure}
\usepackage{dcolumn}
\usepackage{bm}
\usepackage{color}
\usepackage{xcolor}
\usepackage{CJK}
\usepackage{subfigure}
\usepackage{ulem}

\newcommand{\beq}{\begin{equation}}
\newcommand{\eeq}{\end{equation}}
\newcommand{\beqa}{\begin{eqnarray}}
\newcommand{\eeqa}{\end{eqnarray}}

\begin{document}

\title{Nonlinear quantum Rabi model in trapped ions}

\author{Xiao-Hang Cheng}
\affiliation{Department of Physics, Shanghai University, 200444 Shanghai, People's Republic of China}
\affiliation{Department of Physical Chemistry, University of the Basque Country UPV/EHU, Apartado 644, 48080 Bilbao, Spain}

\author{I\~{n}igo Arrazola}
\affiliation{Department of Physical Chemistry, University of the Basque Country UPV/EHU, Apartado 644, 48080 Bilbao, Spain}

\author{Julen S. Pedernales}
\affiliation{Department of Physical Chemistry, University of the Basque Country UPV/EHU, Apartado 644, 48080 Bilbao, Spain}
\affiliation{Institute for Theoretical Physics and IQST, Albert-Einstein-Allee 11, Universit\"at Ulm, D-89069 Ulm, Germany}

\author{Lucas Lamata}
\affiliation{Department of Physical Chemistry, University of the Basque Country UPV/EHU, Apartado 644, 48080 Bilbao, Spain}

\author{Xi Chen}
\affiliation{Department of Physics, Shanghai University, 200444 Shanghai, People's Republic of China}

\author{Enrique Solano}
\affiliation{Department of Physics, Shanghai University, 200444 Shanghai, People's Republic of China}
\affiliation{Department of Physical Chemistry, University of the Basque Country UPV/EHU, Apartado 644, 48080 Bilbao, Spain}
\affiliation{IKERBASQUE, Basque Foundation for Science, Maria Diaz de Haro 3, 48013 Bilbao, Spain}

\begin{abstract}
We study the nonlinear dynamics of trapped-ion models far away from the Lamb-Dicke regime. This nonlinearity induces a blockade on the propagation of quantum information along the Hilbert space of the Jaynes-Cummings and quantum Rabi models. We propose to use this blockade as a resource for the dissipative generation of high-number Fock states. Also, we compare the linear and nonlinear cases of the quantum Rabi model in the ultrastrong and deep strong coupling regimes. Moreover, we propose a scheme to simulate the nonlinear quantum Rabi model in all coupling regimes. This can be done via off-resonant nonlinear red and blue sideband interactions in a single trapped ion, yielding applications as a dynamical quantum filter.
\end{abstract}


\maketitle
\section{Introduction}
Proposed in 1936 by I. I. Rabi~\cite{Rabi}, the most fundamental interaction between a two-level atom and a classical light field, the semiclassical Rabi model, has played an important role in both physics and mathematics~\cite{Rabi80, analytic Rabi}. Under the rotating wave approximation (RWA), its fully quantized form, the quantum Rabi model (QRM), can be reduced into the Jaynes-Cummings model (JCM)~\cite{JCM}, which is analytically solvable~\cite{Wolfgang}. This model describes the basic interaction in trapped ions~\cite{ion review}, superconducting circuits~\cite{superconducting1}, and cavity quantum electrodynamics~\cite{CQED}, when the systems are in the regime where the ratio of coupling strength $g$ and mode frequency $\nu$ is approximately smaller than $0.1$~\cite{structure}. On the other hand, in the ultrastrong coupling (USC) regime, $g/\nu \in (0.1, 1)$~\cite{USC circuits, USC2}, and deep strong coupling (DSC) regime $(g/\nu > 1)$~\cite{DSC, DSC2, spectral}, we have to take the counter rotating term that is neglected in the JCM into account. The QRM is a fruitful physical model with applications in condensed matter, quantum optics, and quantum information processing. In fact, the QRM has been investigated in many contexts, such as quantum phase transitions~\cite{QPT1, ground state, QPT2}, dissipative QRM~\cite{SDE Rabi}, generalized QRM~\cite{GQRM1, GQRM2, GQRM3, GQRM4}, multiparticle QRM~\cite{two qubits Rabi, MultiQRM1, MultiQRM2, MultiQRM3}, and quantum thermodynamics~\cite{heat engine}, among others. Furthermore, proposals and experimental realizations of the QRM in different quantum simulators as optical lattices~\cite{ultracold atoms}, circuit QED~\cite{Rabi circuits}, as well as trapped ions~\cite{ion Rabi, ion Rabi exp, cross cavity Rabi,PueblaNJP} have been put forward. Reference~\cite{ion Rabi} introduced an analog method for the simulation of different regimes of the QRM, otherwise inaccessible to experimentation from first principles. These ideas have been recently demonstrated in the lab \cite{ion Rabi exp}.

As one of the most controllable quantum systems, trapped ions play an important role in diverse proposals for quantum simulations~\cite{ion Rabi, ion Rabi exp, trapped ion3, spin models, QFT, fermion lattice, Holstein model, fermionic and bosonic models, qs dirac, nature, Klein Paradox, majorana1, KihwanMajo, parity, Arrazola2016,Cheng2017}. However, most of these works are based on the condition for the system to be in the Lamb-Dicke (LD) regime. In this regime, the size of the motional wavepacket of the ion is much smaller than the wavelength of the external laser driving, such that the effective coupling between the internal and the vibrational degrees of freedom, generated by the laser field, can be approximated to first order~\cite{ion review}. This condition can also be expressed as $\eta\sqrt{\langle (a+a^{\dag})^2 \rangle}\ll 1$, where $a^{\dag}$($a$) is the creation (annihilation) operator associated to the quantum vibrational mode of the ion on a certain direction $x$ and $\eta=k\sqrt{\hbar/2 M\nu}$ is the LD parameter in this direction, with $k$ the wave number of the external laser field, $M$ the mass of the ion and $\nu$ the frequency of the harmonic potential.

In this article, we study the nonlinear behaviour of a single trapped ion when it is far away from the LD regime. In the past, research beyond the LD regime was mainly focussed on the nonlinear JCM~\cite{Vogel95,MatosFilho96,MatosFilho96_2,Stevens98}, but has also been studied for its implications in laser cooling~\cite{Morigi97,Morigi99,Foster09} or for its possible applications to simulate Frack-Condon physics~\cite{Hu11}. To set up the stage for a subsequent analysis, we first briefly review the JCM and take this as a reference to show the difference with the nonlinear JCM. The appearance of nonlinear terms in the Hamiltonian suppresses the collapses and revivals for a coherent state evolution typical from linear cases.  Later on, we investigate how the nonlinear anti-Jaynes-Cummings model, which appears as the counterpart of nonlinear JCM, can be combined with controlled depolarizing noise, to generate arbitrary $n$-phonon Fock states. Moreover, the latter could in principle be done without a precise control of pulse duration or shape, and without the requirement of a previous high-fidelity preparation of the motional ground state. Furthermore, we propose the quantum simulation of the nonlinear quantum Rabi model by simultaneous off-resonant nonlinear Jaynes-Cummings and anti-Jaynes-Cummings interactions. Finally, we also point out the possibility for the quantum Rabi model to act as a motional state filter.

\section{Jaynes-Cummings Models in Trapped Ions}
The Hamiltonian describing a laser-cooled two-level ion trapped in a harmonic potential and driven by a monochromatic laser field can be expressed as ($\hbar=1$)
\begin{equation}\label{IonHamil}
H=\frac{\omega_0}{2}\sigma_z+\nu a^{\dag}a+\frac{\Omega}{2}\sigma^x[e^{i(\eta(a+a^\dag)-\omega t+\phi)}+{\rm H.c.}],
\end{equation}
where $\omega_0$ is the two-level transition frequency, $\sigma_z,\sigma^x$ are  Pauli matrices associated to this two-level system, $\Omega$ is the Rabi frequency, $\omega$ is the driving laser frequency, and $\phi$ is the phase of the laser field.

In the Lamb-Dicke regime, moving to an interaction picture with respect to $H_0=\frac{\omega_0}{2}\sigma_z+\nu a^{\dag}a$, and after the application of the so-called optical RWA, the Hamiltonian in Eq.(\ref{IonHamil}) can be written as~\cite{ion review}
\begin{equation}\label{LDregime}
H_{\rm int}^{\rm LD}=\frac{\Omega}{2}\sigma^+[1+i\eta(ae^{-i\nu t}+a^\dag e^{i\nu t})]e^{i(\phi-\delta t)}+{\rm H.c.},
\end{equation}
where $\delta=\omega-\omega_0$ is the laser detuning and the condition $\eta \ll1$ allows to keep only zero and first order terms in the expansion of $\exp{[i\eta(a+a^\dag)]}$. When $\delta=-\nu$ and $\Omega\ll \nu$, after applying the vibrational RWA, the dynamics of such a system is described by Jaynes-Cummings Hamiltonian, $H_{\rm JC}=i g (\sigma^+ a - \sigma^-a^{\dag})$, where $g=\eta\Omega/2$ and $\phi=0$. This JCM is analytically solvable and generates population exchange between states $|\!\downarrow,n\rangle \leftrightarrow |\!\uparrow,n\!-\!1\rangle$ with rate $\Omega_{n,n-1}=\eta\Omega\sqrt{n}$. On the other hand, when the detuning is chosen to be $\delta=\nu$, the effective model is instead described by the anti-JCM $H_{\rm aJC}=i g (\sigma^+ a^\dag - \sigma^-a)$, which generates population transfer between states $|\!\downarrow,n\rangle \leftrightarrow |\!\uparrow,n\!+\!1\rangle$ with rate $\Omega_{n,n+1}=\eta\Omega\sqrt{n+1}$.

When the trapped-ion system is beyond the Lamb-Dicke regime, the simplification of the exponential term described above is not justified and Eq.(\ref{LDregime}) reads
\begin{eqnarray}\label{BLDregime}
H_{\rm int}=\frac{\Omega}{2}\sigma^+ e^{i\eta(a^+ e^{i\nu t}+a e^{-i\nu t})-i(\delta t-\phi)}+{\rm H.c.}\label{IntHam}.
\end{eqnarray}
When $\delta=-\nu$ and $\Omega \ll \nu$, after applying the vibrational RWA, the effective Hamiltonian describing the system is given by the nonlinear Jaynes-Cummings model~\cite{Vogel95}, which can be expressed as
\begin{eqnarray}
H_{\rm nJC}=ig[\sigma^+ f_1(\hat{n}) a - \sigma^- a^\dag f_1(\hat{n})],
\end{eqnarray}
where the nonlinear function $f_1$~\cite{Vogel95} is given by
\begin{equation}\label{NLfunc}
f_1(\hat{n})=e^{-\eta^2/2}\sum_{l=0}^{\infty}\frac{(-\eta^2)^l}{l!(l+1)!}a^{\dag l} a^l,
\end{equation}
with $a^{\dag l} a^l=\hat{n}!/(\hat{n}-l)!$. The dynamics of this model can also be solved analytically, and as the linear JCM, yields to population exchange between states $|\!\downarrow,n\rangle \leftrightarrow |\!\uparrow,n\!-\!1\rangle$, but in this case with a rate $\tilde{\Omega}_{n,n-1}= |f_1(n\!-\!1)|\Omega_{n,n-1}=\eta\Omega\sqrt{n} |f_1(n\!-\!1)|$, where $f_1(n)$ corresponds to the value of the diagonal operator $f_1$ evaluated on the Fock state $|n\rangle$, i.e. $f_1(n)\equiv\langle f_1(\hat{n})\rangle_n$.
If the detuning in Eq.(\ref{BLDregime}) is chosen to be $\delta=\nu$, and $\Omega \ll \nu$, then the application of the vibrational RWA yields the nonlinear anti-JCM,
\begin{eqnarray}
H_{\rm naJC}=ig[\sigma^+a^\dag f_1(\hat{n}) - \sigma^- f_1(\hat{n})a ],
\end{eqnarray}
which, as the linear anti-JCM, generates population exchange between states $|\!\downarrow,n\rangle \leftrightarrow |\!\uparrow,n\!+\!1\rangle$ with rate $\tilde{\Omega}_{n,n+1}=|f_1(n)|\Omega_{n,n+1}=\eta\Omega\sqrt{n+1} |f_1(n)|$. The nonlinear function $f_1$ depends on the LD parameter $\eta$ and on the Fock state $| n \rangle$ on which it is acting. The LD regime is then recovered when $\eta\sqrt{n}\ll 1$. In this regime, $|f_1(n)|\approx1$, and thus the dynamics are the ones that correspond to the linear models.
\begin{figure}[]
{\includegraphics[width=1.0 \linewidth]{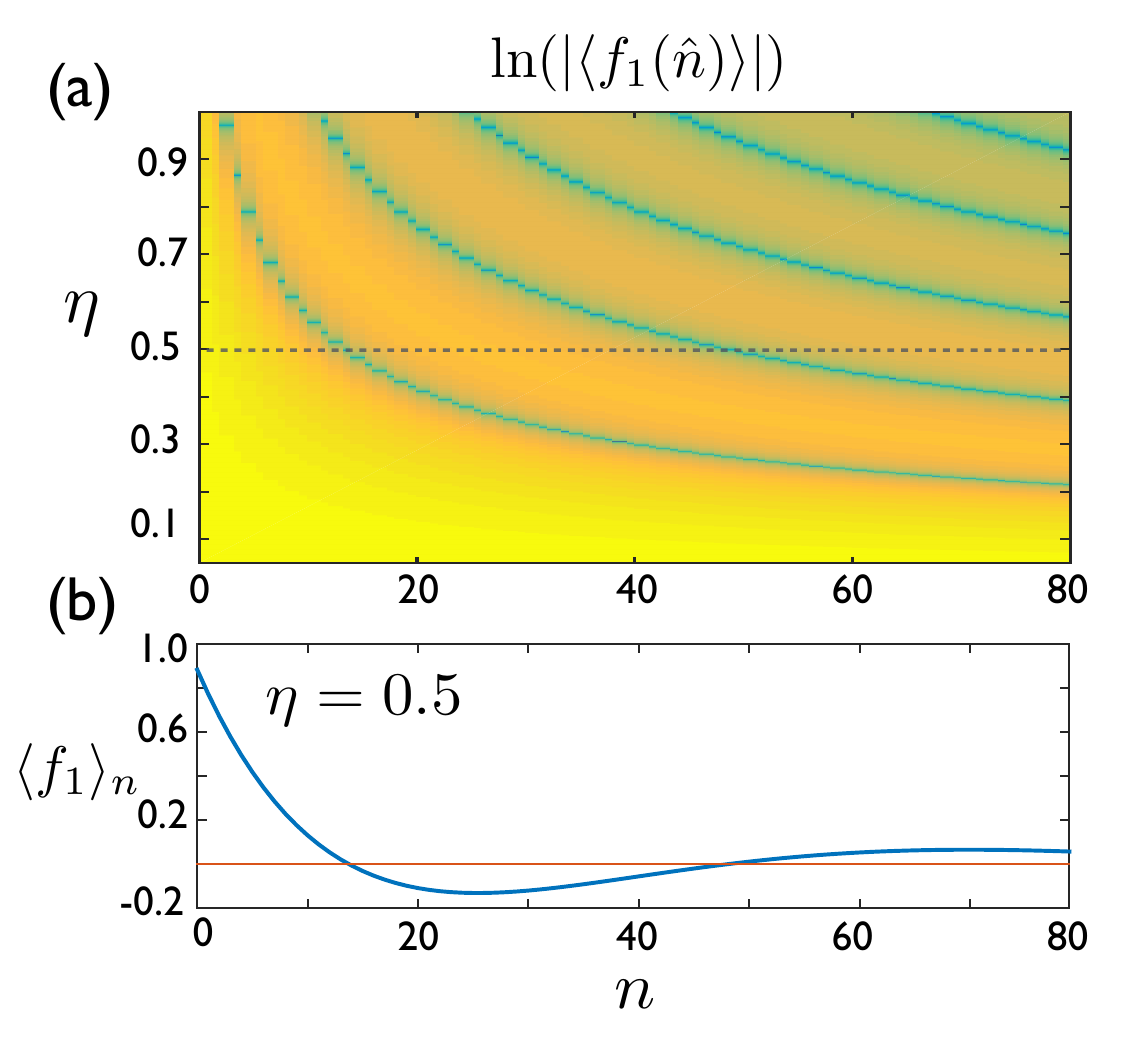}}
\caption{(color online) (a) Logarithm of the absolute value of the operator $f_1(\hat{n})$ evaluated for different Fock states $|n\rangle$ and LD parameters $\eta$. Dark (blue) regions represent cases where $f_1(\hat{n})|n\rangle\approx 0$. (b) Nonlinear function $f_1(n)$ for a fixed value of the LD parameter $\eta=0.5$ (oscillating blue curve). Zero value (horizontal orange line)\label{NLZeros}}
\end{figure}

Beyond the LD regime the nonlinear function $f_1$, which has an oscillatory behaviour both in $n\in\mathbb{N}$ and $\eta\in\mathbb{R}$, needs to be taken into account.  In Fig.~\ref{NLZeros}a, we plot the logarithm of the absolute value of $f_1(n,\eta)$ for different values of $n$ and $\eta$, where the green regions represent lower values of $\log{(|f_1(n,\eta)|)}$, i.e, values for which $f_1\approx 0$. This oscillatory behaviour can also be seen in Fig.~\ref{NLZeros}b where we plot the value of $f_1$ as a function of the Fock state number $n$ for $\eta=0.5$. For this specific case, we can see that the function is close to zero around $n=14$ and $n=48$, meaning that for $\eta=0.5$, the rate of population exchange between $|\!\downarrow,15\rangle \leftrightarrow |\!\uparrow,14\rangle$ and $|\!\downarrow,49\rangle \leftrightarrow |\!\uparrow,48\rangle$ states on the nonlinear JCM will  vanish. The same will happen to the exchange rate between $|\!\downarrow,14\rangle \leftrightarrow |\!\uparrow,15\rangle$ and $|\!\downarrow,48\rangle \leftrightarrow |\!\uparrow,49\rangle$ states for the nonlinear anti-JCM.

\begin{figure}[]
{\includegraphics[width=1.0 \linewidth]{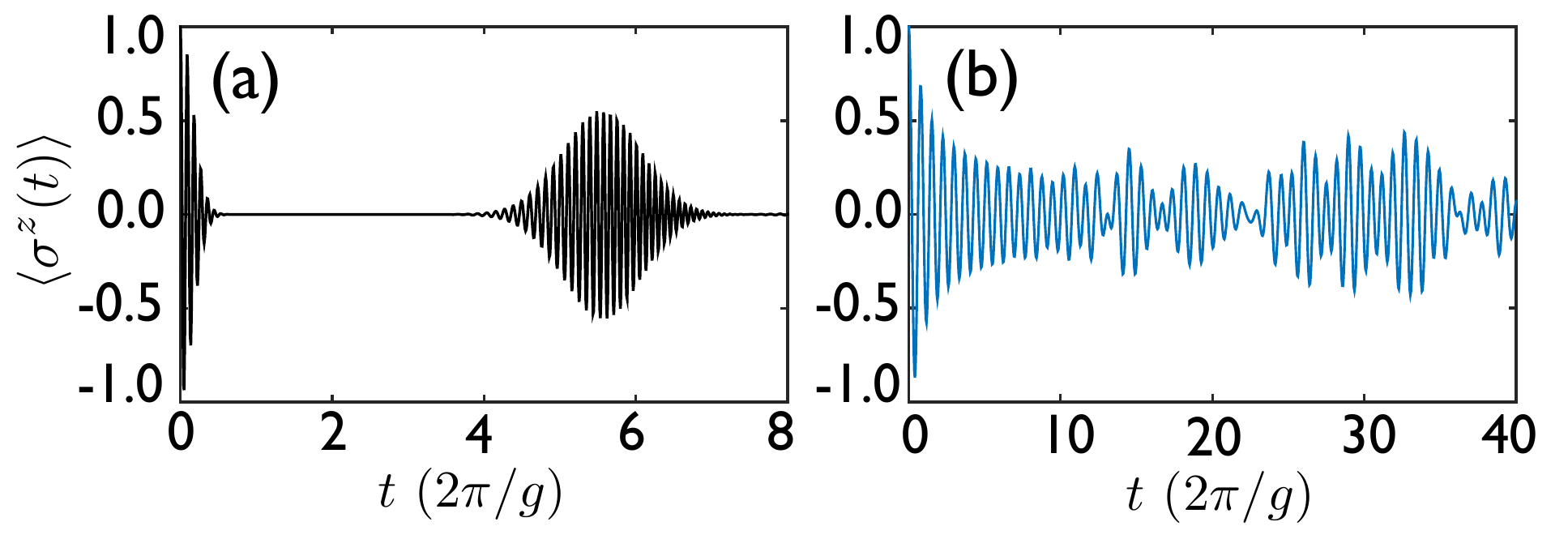}}
\caption{(color online) Average value of $\sigma_z$ operator versus time for a coherent initial state $|\alpha=\sqrt{30}\rangle$ after (a) linear JC and (b) nonlinear JC evolution, both with the same coupling strength $g$ and $\eta=0.5$ for the nonlinear case. As shown in (a), there exists an approximate collapse and subsequent revival in the JCM dynamics, while for the nonlinear JCM this is not the case. \label{JaynesCollapse}}
\end{figure}

We observe approximate collapses and revivals for an initial coherent state with an average number of photons of $|\alpha|^2=30$ by evolving with the JCM, as shown in Ref.~\cite{CollapseRevival}, see Fig.~\ref{JaynesCollapse}a. Here, we plot $\langle\sigma^z(t)\rangle=\langle \psi(t)|\sigma^z|\psi(t)\rangle$ for a state that evolves according to the JCM. Comparing the same case for the nonlinear JCM with $\eta=0.5$, as depicted in Fig.~\ref{JaynesCollapse}b, we appreciate that in the latter case the collapses and revivals vanish, and the dynamics is more irregular. This can seem natural given that the phenomenon of revival takes place whenever the most significant components of the quantum state, after some evolution time, turn out to oscillate in phase again, which may be more unlikely if the dynamics is nonlinear. Notice that we let the case of the nonlinear JCM evolve for a longer time, since the nonlinear function $f_1$ effectively slows down the evolution.

\section{Fock State generation with Dissipative nonlinear anti-JCM}

In this section we study the possibility of using the dynamics of the nonlinear anti-Jaynes-Cummings model introduced in the previous section to, along with depolarizing noise, generate high-number Fock states in a dissipative manner. In particular, the depolarizing noise that we consider corresponds to the spontaneous relaxation of the internal two-level system of the ion. Such a dissipative process, combined with the dynamics of the JCM in the LD regime (linear JCM), is routinely exploited in trapped-ion setups for the implementation of sideband cooling techniques. It is noteworthy to mention that the effect of nonlinearities on sideband cooling protocols, which arise when outside the LD regime, have also been a matter of study~\cite{cooling1, beyond LD}.

Our method works as follows: we start in the ground state of both the motional and the internal degrees of freedom $|\!\downarrow,0\rangle$ (as we will show later, our protocol works as well when we are outside the motional ground state, as long as the population of Fock states higher than the target Fock state is negligible). Acting with the nonlinear anti-JC Hamiltonian we induce a population transfer from state $|\!\downarrow,0\rangle$ to state $|\!\uparrow,1\rangle$, while at the same time, the depolarizing noise transfers population from $|\!\uparrow, 1\rangle$ to $|\!\downarrow, 1\rangle$. The simultaneous action of both processes will ``heat" the motional state, progressively transferring the population of the system from one Fock state to the next one. Eventually, all the population will be accumulated in state $|\!\downarrow,n\rangle$, where a blockade of the propagation of population through the chain of Fock states occurs, if $f_1(n)=0$, as the transfer rate between states $|\!\downarrow,n\rangle$ and $|\!\uparrow,n+1\rangle$ vanishes, $\tilde{\Omega}_{n,n+1}=0$. We point out that the condition $f_1(n)=0$ can always be achieved by tuning the LD parameter to a suitable value, i.e. for every Fock state $|n\rangle$, where $n>0$ there exists a value of the LD parameter $\eta$ for which $f_1(n,\eta)=0$. As an example, we choose the LD parameter $\eta=0.4518$, for which $f_1(17)=0$, and simulate our protocol using the master equation
\begin{eqnarray}\label{LabMasterEq}
\dot{\rho}=&&-i[H_{\rm naJC},\rho]+\Gamma_{m} L(\sigma^{-})\rho,
\end{eqnarray}
where  $\Gamma_{m}=2g$ is the decay rate of the internal state, and the Lindblad superoperator acts on $\rho$ as ${L(\hat{X})\rho=(2\hat{X}\rho \hat{X}^{\dag}-\hat{X}^{\dag} \hat{X}\rho-\rho \hat{X}^{\dag} \hat{X})/2}$.
\begin{figure}[h]
{\includegraphics[width=1 \linewidth]{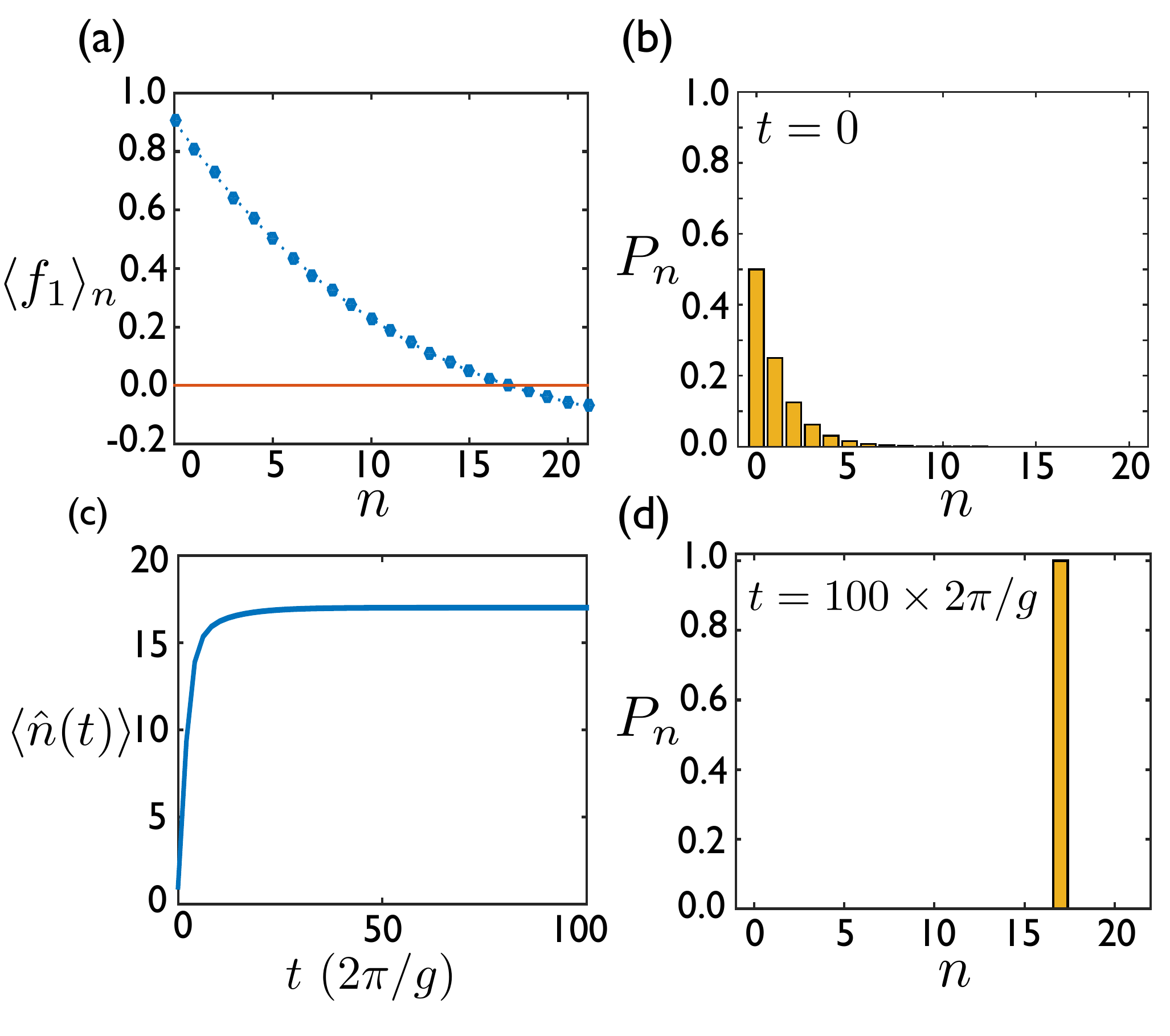}}
\caption{(color online) (a) The nonlinear function $f_1$ evaluated at different Fock states $n$, for the case of $\eta=0.4518$ (decreasing blue curve). Zero value (horizontal orange line). For this value of the LD parameter, $f_1|17\rangle=0$. (b) Phonon statistics of the initial thermal state with $\langle n \rangle=1$ (c) Time evolution of the average value of the number operator $\hat{n}$ starting from the state in (b) and following the evolution for the preparation of Fock state $| 17 \rangle$, that is during a nonlinear anti-JCM with spontaneous decay of the two-level system. (d) Phonon statistics at the end of the protocol, $t=100\times2\pi/g$, with all the population concentrated in Fock state $|17\rangle$. \label{ajcdecay}}
\end{figure}

In Fig.~\ref{ajcdecay} we numerically show how our protocol is able to generate the motional Fock state $ |17 \rangle$, starting from a thermal state $\rho_T=\sum_{k=0}^{\infty} \frac{\langle n\rangle^k}{(\langle n\rangle+1)^{k+1}}|k\rangle\langle k|$, with $\langle n\rangle=1$. In other words, one can obtain large final Fock states starting from an imperfectly cooled motional state, by a suitable tunning of the LD parameter. As an advantage of our method compared to previous approaches~\cite{Meekhof96}, we do not need a fine control over the Rabi frequencies or pulse durations, given that the whole wavefunction, for an arbitrary initial state with motional components smaller than $n$, will converge to the target Fock state $|n\rangle$. We want to point out that this protocol relies only on the precision to which the LD parameter can be set, which in turn depends on the precision to which the wave number $k$ and the trap frequency $\nu$ can be controlled. These parameters enjoy a great stability in trapped-ion setups~\cite{Johnson16}, and therefore we deem the generation of high-number Fock states as a promising application of the nonlinear anti-JCM dynamics.

\section{Nonlinear Quantum Rabi Model}

Here we propose to implement the nonlinear quantum Rabi model (NQRM) in all its parameter regimes via the use of the Hamiltonian in Eq.(\ref{IntHam}). We consider off-resonant first-order red- and blue-sideband drivings with the same coupling $\Omega$ and corresponding detunings $\delta_r$, $\delta_b$. The interaction Hamiltonian after the optical RWA reads~\cite{ion review, ion Rabi},
\begin{eqnarray}
H_{\rm int} = \sum\limits_{n=r,b}\frac{\Omega}{2}\sigma^+e^{i\eta(a^{\dag}e^{i \nu t}+a e^{-i \nu t})}e^{-i(\delta_n t-\phi_n)}+{\rm H.c.},
\end{eqnarray}
where $\omega_r=\omega_0-\nu+\delta_r$ and $\omega_b=\omega_0+\nu+\delta_b$, with $\delta_r,\delta_b\ll \nu \ll \omega_0$ and $\Omega \ll \nu$. We consider the system beyond the Lamb-Dicke regime and set the laser field phases to $\phi_{r,b}=0$. If we invoke the vibrational RWA, i.e. neglect terms that rotate with frequencies in the order of $\nu$, the remaining terms read
\begin{equation}
H_{\rm int}=ig\sigma^+\big(f_1ae^{-i \delta_r t}+a^{\dag}f_1e^{-i \delta_b t}\big)+{\rm H.c.},
\end{equation}
where  $g=\eta\Omega/2$ and  $f_1\equiv f_1(\hat{n})$ was introduced in Eq.~(\ref{NLfunc}). The latter corresponds to an interaction picture Hamiltonian of the NQRM with respect to the free Hamiltonian $H_0=\frac{1}{4}(\delta_b+\delta_r)\sigma_z +\frac{1}{2}(\delta_b-\delta_r)a^\dag a$. Therefore, undoing the interaction picture transformation, we have
\begin{equation}\label{NQRM}
H_{\rm nQRM}=\frac{\omega_0^{\rm R}}{2}\sigma_z+\omega^{\rm R} a^{\dag}a+i g (\sigma^+ - \sigma^-)(f_1a+a^{\dag}f_1),
\end{equation}
where $\omega_0^{\rm R}=-\frac{1}{2}(\delta_r+\delta_b)$ and $\omega^{\rm R}=\frac{1}{2}(\delta_r-\delta_b)$.  Equation~(\ref{NQRM}) represents the general form of the NQRM, where $\omega_0^{\rm R}$ is the level splitting of the simulated two level system, $\omega^{\rm R}$ is the frequency of the simulated bosonic mode and $g$ is the coupling strength between them, which in turn will be modulated by the nonlinear function $f_1(\hat{n},\eta)$. The different regimes of the NQRM will be characterized by the relation among these four parameters. First, in the LD regime or $\eta\sqrt{\langle (a+a^{\dag})^2 \rangle}\ll 1$, Eq.~(\ref{NQRM}) can be approximated to the linear QRM~\cite{ion Rabi}. Beyond the LD regime, in a parameter regime where $|\omega^{\rm R}-\omega_0^{\rm R}|\ll g\ll|\omega^{\rm R}+\omega_0^{\rm R}|$, the RWA can be applied. This would imply neglecting terms that rotate at frequency $\omega^{\rm R}+\omega_0^{\rm R}$ in an interaction picture with respect to $H_0$, leading to the nonlinear JCM studied previously in this article. On the other hand, the nonlinear anti-JCM would be recovered in a regime where $|\omega^{\rm R}+\omega_0^{\rm R}|\ll g\ll|\omega^{\rm R}-\omega_0^{\rm R}|$. It is worth mentioning that the latter is only possible if the frequency of the two-level system and the frequency of the mode have opposite sign. The USC and DSC regimes are defined as $0.1\lesssim g/\omega^{\rm R} \lesssim 1$ and $g/\omega^{\rm R}\gtrsim1$ respectively, and in these regimes the RWA does not hold anymore.

\begin{figure}[]
{\includegraphics[width=1 \linewidth]{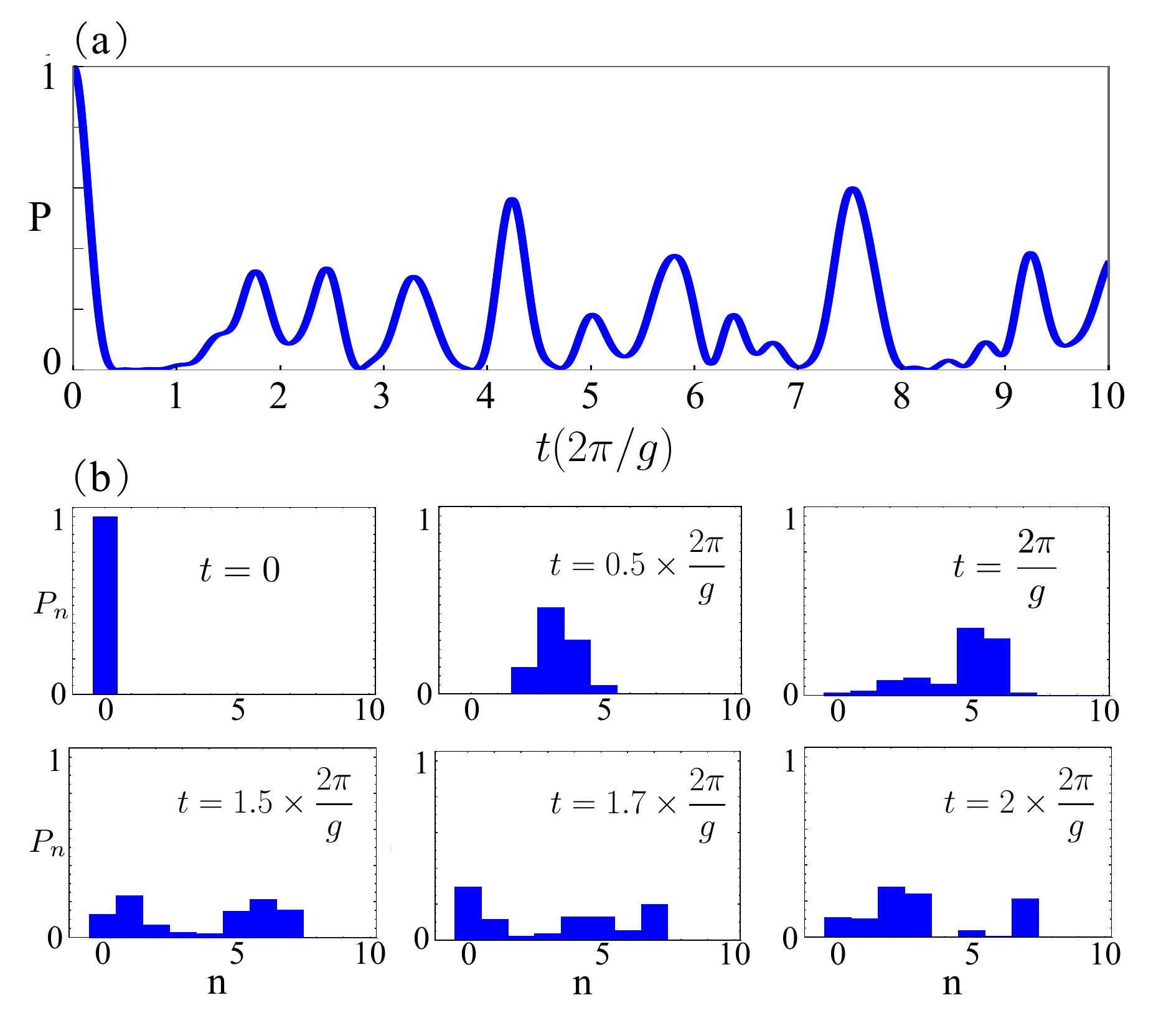}}
\caption{(color online) (a) Fidelity with respect to the initial state $P(t)=|\langle\psi_0|\psi(t)\rangle|^2$ versus time. As initial state we choose $|0, \rm g\rangle$ and the evolution occurs under the NQRM with LD parameter $\eta=0.67898$, where $f_1|7\rangle=0$, $g/\omega^{\rm R}=4$ and $\omega_0^{\rm R}=0$. (b) Phonon statistics at different times for the NQRM evolved from the initial state $|0, \rm g\rangle$. The propagation of population through Fock states stops at $|n=7\rangle$, with Fock states of $n>7$ never getting populated. \label{NonlinearFockProSta}}
\end{figure}

As an example, here we investigate the NQRM in the DSC regime with initial Fock state $|0, \rm g\rangle$, where $|0\rangle$ is the ground-state of the bosonic mode, and $|\rm g \rangle$ stands for the ground state of the effective two-level system. In Fig.~\ref{NonlinearFockProSta}, we study the case for $\eta=0.67898$, where $f_1|7\rangle=0$, $g/\omega^{\rm R}=4$ and $\omega_0^{\rm R}=0$. More specifically, a quantum simulation of the model in this regime can be achieved with the following detunings and Rabi frequency: $\delta_r=2\pi\times11.31$kHz, $\delta_b=-2\pi\times11.31$kHz, $g=2\pi\times45.24$kHz and $\Omega=2\pi\times 133.26$kHz. In Ref.~\cite{DSC}, it was shown that the linear QRM shows collapses and revivals and a round trip of the phonon-number wavepacket along the chain of Fock states, when in the DSC regime. Here, we observe that in the nonlinear case, Fig.~\ref{NonlinearFockProSta}, collapses and revivals do not present the same clear structure, having a more irregular evolution. Most interestingly, the system dynamics never surpasses Fock state $|n\rangle$, for which $f_1(n)=0$.  Regarding the simulated regime of the nonlinear QRM, we point out that the nonlinear term also contributes to the coupling strength. Therefore, to keep the NQRM in the DSC regime, the ratio $g/\omega^{\rm R}$ should be larger than that for the linear QRM since $f_1(n)< 1$ always. Summarizing, our result illustrates that the Hilbert space is effectively divided into two subspaces by the NQRM, namely those spanned by Fock states below and above Fock state $| n \rangle$.  We denote the Fock number $n$, where $f_1|n\rangle=0$, as ``the barrier'' of the NQRM.
\begin{figure}[]
{\includegraphics[width=1 \linewidth]{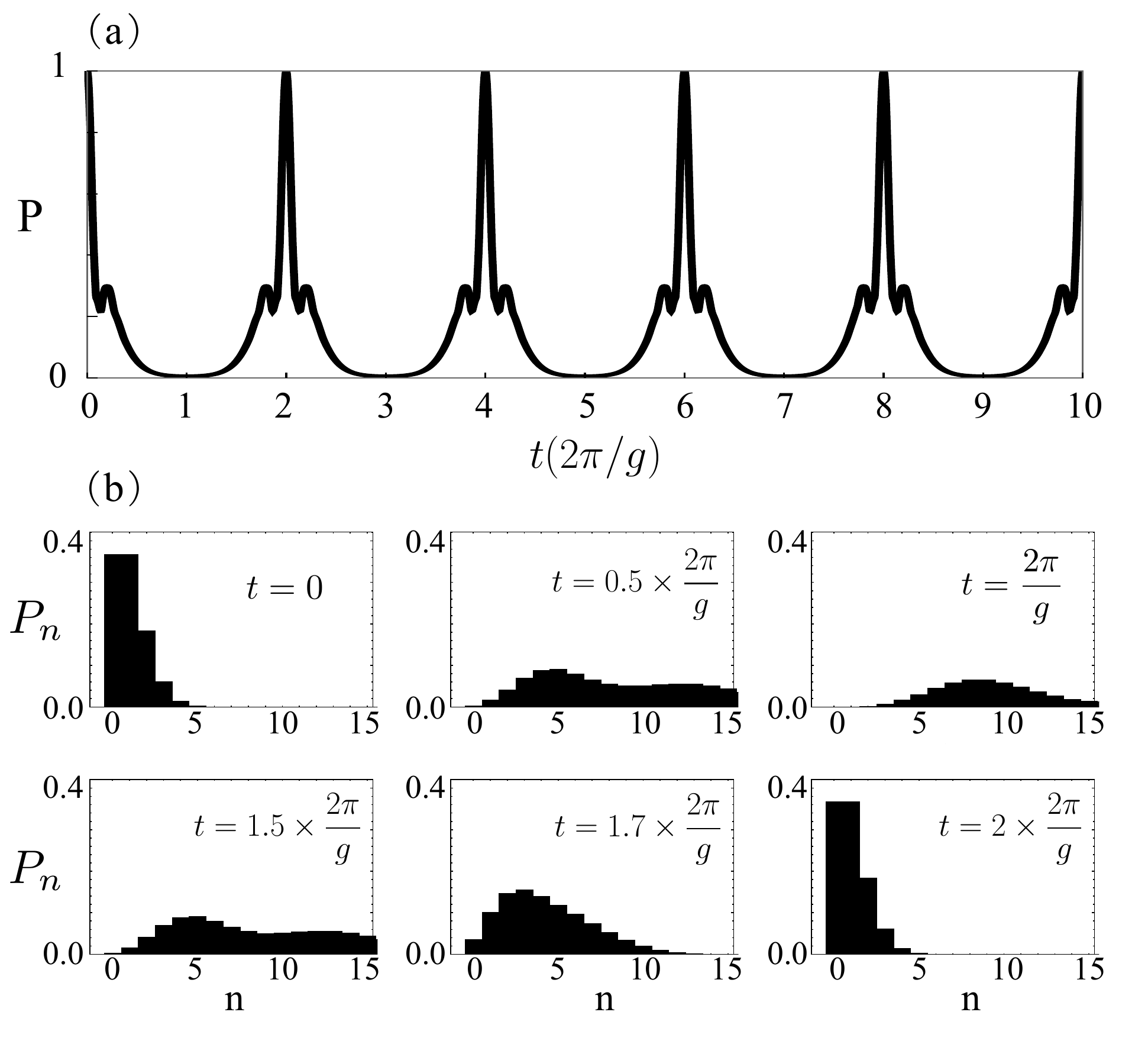}}
\caption{(color online) (a) Overlap of the instantaneous state with the initial state $P(t)=|\langle\psi_0|\psi(t)\rangle|^2$ versus time, for a coherent initial state $|\alpha\!=\!1,\rm g\rangle$ evolving under the linear QRM. Collapses and revivals are observed, as expected in the DSC regime of the linear QRM. (b) Phonon statistics at different times, where we see the round trip of a phonon number wavepacket.  \label{LinearCohProSta}}
\end{figure}
\begin{figure}[]
{\includegraphics[width=1 \linewidth]{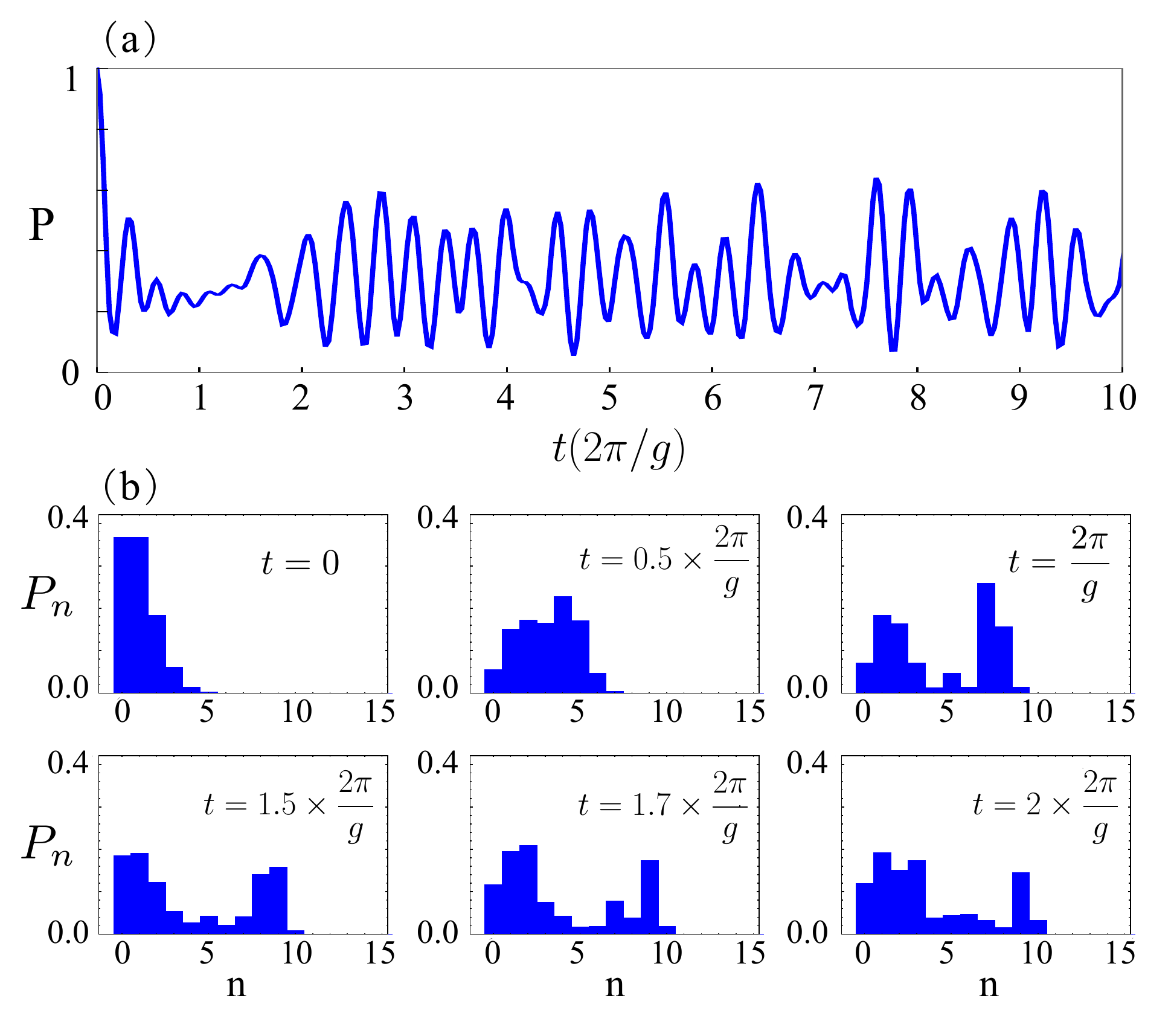}}
\caption{(color online) (a) Overlap with the initial state $P(t)=|\langle\psi_0|\psi(t)\rangle|^2$ versus time, for initial state $|\alpha\!=\!1,\rm g\rangle$ evolving under the NQRM with LD parameter $\eta=0.57838$, where $f_1|10\rangle=0$, $g/\omega^{\rm R}=3.7$ and $\omega_0^{\rm R}=0$. (b) Phonon statistics at different times for the NQRM evolved from the initial state $|\alpha\!=\!1,\rm g\rangle$. The Fock state $|10\rangle$ is never surpassed because $f_1|10\rangle=0$. \label{NonlinearCohProSta}}
\end{figure}
To benchmark the effect of the barrier, we also provide simulations starting from an initial coherent state with $\alpha=1$ whose average phonon number is $\langle n \rangle=|\alpha|^2=1$, and make the comparison between the QRM and the NQRM in the DSC regime. For the parameter regime $g/\omega^{\rm R}=2$ and $\omega_0^{\rm R}=0$, the fidelity with respect to the initial coherent state in the linear QRM performs periodic collapses and full revivals as it can be seen in Fig.~\ref{LinearCohProSta}(a). In Fig.~\ref{LinearCohProSta}(b), we observe a round trip of the phonon-number wave packet, similarly to what was shown in Ref.~\cite{DSC} for the case of the linear QRM starting from a Fock state. The NQRM, on the other hand, has an associated dynamics that is aperiodic and more irregular, as shown in Fig.~\ref{NonlinearCohProSta}, and never crosses the motional barrier produced by the corresponding $f_1(n)=0$. Therefore, it can be employed as a motional filter, which is determined by the location of the barrier with respect to the initial state distribution. Here, by filter we mean that the population of Fock states above a given threshold can be prevented. For the simulation we choose the LD parameter $\eta=0.57838$ for which $f_1|10\rangle=0$, which is far from the center of the distribution of the initial coherent state, as well as most of its width. The simulated parameter regime corresponds to the DSC regime with $g/\omega^{\rm R}=3.7$ and $\omega_0^{\rm R}=0$. This case could also be simulated with trapped ions with detunings of $\delta_r=2\pi\times11.31$kHz and $\delta_b=-2\pi\times11.31$kHz, and a Rabi frequency of $\Omega=2\pi\times 133.26$kHz. As for the corresponding case with initial Fock state $|0,\rm g\rangle$, the evolution of the NQRM in the coherent state case, depicted in Fig.~\ref{NonlinearCohProSta}, never exceeds the barrier.

\section{Conclusions}

We have proposed the implementation of nonlinear QRMs in arbitrary coupling regimes, with trapped-ion analog quantum simulators. The nonlinear term that appears in our model is characteristic of the region beyond the Lamb-Dicke regime.  This nonlinear term causes the blockade of motional propagation at $|n\rangle$, whenever $f_1(\hat{n})|n\rangle=0$. In order to compare our models with standard linear quantum Rabi models, we have plotted the evolution of the population of the internal degrees of freedom of the ion evolving under the linear JCM and the nonlinear JCM, and observe that for the latter the collapses and revivals disappear. Also, we have proposed a method for generating large Fock states in a dissipative manner, making use of the nonlinear anti-JCM and the spontaneous decay of the two-level system. Finally, we have studied the dynamics of the linear and nonlinear full QRM on the DSC regime and notice that the nonlinear case can act as a motional filter. Our work sheds light on the field of nonlinear QRMs implemented with trapped ions, and suggests plausible applications.

{\it Acknowledgements.}---The authors acknowledge support from NSFC (11474193), the Shuguang Program (14SG35), the Program for Eastern Scholar, the Basque Government with PhD grant PRE-2015-1-0394 and grant IT986-16, Ram\'{o}n y Cajal Grant RYC-2012-11391, MINECO/FEDER FIS2015-69983-P, and Chinese Scholarship Council (201506890077).

\end{document}